\documentclass[twocolumn,showpacs,preprintnumbers,prb,amsmath,amssymb]{revtex4}
\usepackage{graphicx}% Include figure files
\usepackage{dcolumn}% Align table columns on decimal point
\usepackage{bm}% bold math
\begin{document}

\title{Path Integral Monte Carlo study of phonons in the bcc phase of $^4$He
  }

\author{ V.Sorkin }
\email{phsorkin@techunix.technion.ac.il }
\homepage{ http://phycomp.technion.ac.il/~phsorkin/index.html}
\author{ E. Polturak}
\author{ Joan Adler}
\affiliation{ Physics Department, Technion - Israel Institute of Technology, Haifa,
Israel, 32000 }

\date{\today}

\begin{abstract}

                Using  Path Integral Monte Carlo and the Maximum Entropy method, we calculate
                the dynamic structure factor of solid $^4$He in the bcc phase at a finite temperature
                 of T = 1.6 K and a molar volume of 21 cm$^3$.  Both the single-phonon contribution to the
                 dynamic structure factor and the total dynamic
                 structure factor are evaluated. From the dynamic structure
                 factor, we obtain the phonon dispersion relations
                 along the main crystalline directions, [001], [011] and
                 [111]. We calculate both the longitudinal and transverse
                 phonon branches. For the latter, no previous simulations exist. We discuss
                 the differences between dispersion relations
                 resulting from the single-phonon part vs. the total dynamic structure factor.
                  In addition, we evaluate the formation
                 energy of a vacancy.

\end{abstract}
% PACS, the Physics and
% Classification Scheme.

% Classification Scheme.
%Use showkeys class option if keyword
                              %display desired
\maketitle

\section{\label{sec:level1}  Introduction}

Solid helium, the well-known example of a quantum solid, continues
to be a subject of interest to theorists and experimentalists
alike. It is characterized by
large zero-point motion and significant short-range correlation of
its atoms. These effects make the theoretical description of the
solid very difficult. The self - consistent phonon (SCP)
method,\cite{Glyde,Horner} which has been
developed over the years to treat this
problem, takes into account the high anharmonicity and short-range
correlations in order to calculate the dynamical properties of
solid helium.  The predictions of the SCP agree well with experiment
in the fcc and hcp solid phases. In the
low-density bcc phase, the agreement between the theory and
experiment is less satisfactory, in particular regarding the
transverse phonons along the [110] direction.\cite{Gov} The SCP
theory is a variational perturbative theory, and is implemented
at zero-temperature.\cite{Glyde} As a complementary
approach to the SCP, numerical simulations have been performed over the
years. Boninsegni and Ceperley\cite{spect} used Path Integral Monte Carlo
(PIMC) to calculate the phonon spectrum of liquid $^4$He at finite
temperature. Galli and Reatto
\cite{Reatto} used the Shadow Wave Function approach to obtain the
spectra of longitudinal phonons in hcp and bcc $^4$He at zero
temperature.

Recently, interest in the properties of quantum solids has been
revived, following reports indicating the possible existence of a
"supersolid" in the hcp phase\cite{Kim}. In addition, an
optic-like excitation branch was recently discovered in the bcc
phase by Markovich et al,~\cite{Emil,Tuvi} (in a mono-atomic
cubic solid, one should observe only 3 acoustic phonon branches).
These results indicate that the physics of solid helium is not yet
entirely understood. Gov et. al.\cite{Gov} proposed that the new
excitation branch is a result of the coupling of transverse
phonons to additional degrees of freedom, unique to a quantum
solid.

In order to reexamine some of these issues using an alternative
approach, we decided to study the excitations in bcc solid helium
$^4$He, by performing Quantum Monte Carlo (QMC) simulations at a
finite temperature. We use Path Integral Monte Carlo
(PIMC)\cite{Ceperley_RMP,Book,Bernu}, which is a non-perturbative
numerical method, that allows, in principle, simulations of
quantum systems without any assumptions beyond the Schr$ \rm \ddot
o$dinger equation. The two body interatomic He-He potential
\cite{Aziz} is the only input for the PIMC simulations. In our
study the Universal Path Integral (UPI) code of
Ceperley\cite{Ceperley_RMP} was adapted to calculate the phonon
branches at finite temperature.

The novel features of our study include the calculation of the
transverse phonon branches of bcc $^4$He at a finite temperature
of 1.6 K where this phase is stable. Transverse phonons are of
particular interest due to their possible relation with the new
optic-like excitation\cite{Gov}. For longitudinal phonons, we
observe a difference between dispersion relations resulting from
the single-phonon part of dynamic structure factor and the total
structure factor. This difference becomes significant at large
wavevectors. Finally, we repeated our calculations of the
longitudinal phonon spectra in the presence of point defects and
evaluate the formation energy of a vacancy at a constant density
and at a constant volume. We describe details of our simulations
in Sec. II. In Sec. III we present the results of our
calculations, and summarize them in Sec. IV.

\section{\label{sec:level1}  Method }
\subsection{\label{sec:level1} Theory }
The PIMC method used in our simulations is  based on the formulation
of quantum mechanics in terms of path integrals. It has been described in
detail by Ceperley ~\cite{Ceperley_RMP}. The method involves
mapping of the quantum system of particles onto a classical model of
interacting ``ring polymers'', whose elements, ``beads'' or
``time-slices'', are connected by "springs". The method provides a
direct statistical evaluation of quantum canonical averages. In
addition to static properties of the system, dynamical properties
can be also extracted from PIMC simulations.\cite{Ceperley_RMP}

The object of this study is the phonon spectrum, which can be
extracted from the dynamic structure factor, $S({ \bf Q},\omega)$.
We would like to express $S({ \bf Q},\omega)$ in terms of phonon
operators~\cite{Glyde}. The definition of $S({ \bf Q},\omega)$ in
terms of density fluctuations is
\begin{equation}
\label{struct_fun}
S({ \bf Q}, \omega) =  \frac{1}{2 \pi n} \int_{-\infty}^{+\infty} dt e^{i\omega t}<\rho_{ { \bf Q}}(t) \rho_{-{ \bf Q}}(0) >,
\end{equation}
where $\hbar { \bf Q}$ and $\hbar \omega$ are the momentum and
energy ( we take $\hbar = 1$), $\rho_{ { \bf Q}}$ is the
Fourier transform of the density of the solid, and $n$ is the
number density. $S({ \bf Q},\omega)$ is usually expressed in terms
of phonons, by writing $S({ \bf Q},\omega)$ as a sum of terms
involving the excitation of a single phonon, $S_1({ \bf
Q},\omega)$, a pair of phonons, $S_2({ \bf Q},\omega)$ and higher
order terms which also include interference between
different terms.~\cite{Glyde,Horner} In most of our simulations we
calculated the $S_1({ \bf Q},\omega)$ term. Some calculations of
$S({ \bf Q},\omega)$ were also performed, and will be discussed
below.

Taking the instantaneous position ${\bf r}(l,t)$ of atom $l$ as
the lattice point ${ \bf R}_l$ plus a displacement ${ \bf u}(l,t)
= {\bf r}(l,t) - {\bf R}(l,t) $, we rewrite $S({ \bf Q}, \omega)$
in terms of these displacements. The one-phonon contribution is
then given by~\cite{Glyde}
\begin{equation}
 S_1({\bf Q},\omega) =
 d^2 ({ \bf Q}) \sum_l e^{  i{ \bf Q} ({ \bf R}_l - {\bf R}_0)  }
 \left< [{ \bf Q} {\bf u}(l,t) ][{\bf Q} {\bf u}(l,0) ]  \right>,
\end{equation}
where $d({ \bf Q}) = <\exp (-\frac{1}{2}({\bf u }{ \bf Q})^2 )> $ is the Debye-Waller factor.
The displacement ${ \bf u}(l,t) $ can be expressed using the
phonon operators $A_{{ \bf q}, \lambda}(t) $~\cite{Born,Bruesch}
\begin{equation}
{ \bf u}(l,t)  = \sum_{{ \bf q},\lambda} A_{{ \bf q}, \lambda}(t)\exp\left(-i{ \bf q} { \bf R_l}  \right) { \bf \hat e_{\lambda}},
\end{equation}
where ${ \bf q} $ is the phonon wave-vector, $\lambda$ is the
phonon branch index, and $\hat e_{\lambda} $ are polarization
vectors, chosen along the directions [001],[011] and [111]. Using
$A_{{ \bf q}, \lambda}(t)$, the one-phonon term $S_1({ \bf Q},
\omega)$  for a specific phonon branch is rewritten
as~\cite{Glyde}
\begin{equation}
S_1 = \sum_{{ \bf q},\lambda } \int_{-\infty}^{+\infty} < A_{{ \bf q}, \lambda}(t) A_{-{ \bf q}, \lambda}(0) >  \Delta_{{ \bf Q},{ \bf q} -{\bf G}} d^2 [{ \bf Q}{\bf \hat e_{\lambda}}]^2 e^{i\omega t}dt,
\end{equation}
where $\Delta_{{ \bf Q},{ \bf q} -{\bf G}}$ is the delta function,
and ${\bf G}$ is a reciprocal lattice vector. We use ${ \bf Q}$,
which lies inside the first Brillouin zone and parallel
to one of $e_{\lambda}$.
Therefore, $S_1({
\bf Q},\omega) = S_1({
\bf q},\omega)$, and $S_1({\bf q},\omega)$ is given by
\begin{equation}
S_1 = \sum_{\lambda } S_{1,\lambda} = \sum_{\lambda }\int_{-\infty}^{+\infty} e^{i\omega t}F_{{ \bf q}, \lambda}(t) dt,
\end{equation}
and
\begin{equation}
\label{fkt}
F_{{ \bf q}, \lambda}(t) = \int_{-\infty}^{+\infty} e^{-i\omega t} S_{1,\lambda}({ \bf q,\omega})d\omega
\end{equation}
is the intermediate scattering function.

We cannot  directly follow the dynamics of helium atoms in real
time using the Quantum Monte Carlo method. However, we can
extract information about the dynamics by means of the analytical
continuation of $F_{{ \bf q}, \lambda}(t)$ to the complex plane~\cite{spect} $t
\rightarrow i\tau$. Using imaginary-time, we obtain

\begin{equation}
\label{upi}
F_{{ \bf q} , \lambda}(\tau) =  \int_{0}^{+\infty}  S_{1,\lambda }({ \bf q} ,\omega)
 \left(  e^{-\omega \tau} + e^{-\omega(\beta -  \tau)}  \right) d\omega
\end{equation}
where $ F_{{ \bf q}, \lambda}(\tau)$ is the intermediate
scattering function, and $\beta = 1/kT$.

In our simulations, we sampled the displacement ${ \bf u}(l,\tau)$
for each ``time-slice'' $\tau$ of the $l$-th atom represented
by a "ring polymer", and calculated $A_{{ \bf q}, \lambda}(\tau)$ by
performing spatial Fourier transformation
\begin{equation}
\label{ag}
 A_{{ \bf q}, \lambda}(\tau) = \sum_{l} {\bf \hat e_{\lambda}} { \bf u}(l,\tau)   \exp\left( i{ \bf q} { \bf R}_l  \right)
\end{equation}
Using (\ref{ag}), $F_{\bf q}$ is obtained as a quantum canonical average
of the product of the phonon operator $< A_{{ \bf q}, \lambda}(\tau) A_{-{ \bf q},
\lambda}(0)>$ in equilibrium.

In order to calculate $S_{1,\lambda}({ \bf q},\omega) $ from
Eq.(\ref{upi}), we need to perform an inverse Laplace
transformation. Performing this inversion is a difficult numerical
problem,~\cite{Ceperley_RMP,MaxEnt} because of the inherent
statistical uncertainty of noisy PIMC data. The noise rules out an
unambiguous reconstruction of the $S_{1,\lambda}({ \bf q},\omega) $.
The best route to circumvent this problem is to apply
the Maximum Entropy (MaxEnt) ~\cite{MaxEnt,Silver} method that
makes the Laplace inversion better conditioned.

The MaxEnt method yields a dynamic scattering function,  $S_{1,\lambda}({ \bf q},\omega) $
 which satisfies Eq. (7) and at the same time
maximizes the conditional probability imposed by our knowledge of
the system. This can be done if some properties of
$S_{1,\lambda}({ \bf q},\omega)$ are known. For example,
the dynamic scattering factor is a non - negative
function, and has certain asymptotic behavior at small and large
$\omega$. In the MaxEnt method the probability to observe of a
given dynamic scattering function is given by
\begin{equation}
\label{me} P(S_{1,\lambda}({\bf Q},\omega)|< F_{{ \bf q},
\lambda}(\tau)>) \sim \exp\left(-\frac{1}{2}\chi^2+\alpha
S_{ent}  \right)
\end{equation}
where  $P$ is the probability to observe $S_{1,\lambda}({\bf Q},\omega)$
for given set of sampled $< F_{{ \bf q},
\lambda}(\tau)>$, $\chi^2$ is the likelihood,
$\alpha$ is a parameter and $\text S_{ent}$ is the entropy~\cite{MaxEnt}. To
simplify our notation, we use $S_1(\omega)$ below to denote by  the
one-phonon dynamic structure  $S_{1,\lambda}({ \bf q},\omega) $ for
a given ${ \bf q}$ and $\lambda$, and omit the explicit dependence
on ${ \bf q}$ and $\lambda$. Similarly, we replace $ F_{{ \bf q},
\lambda}(\tau)$ by $ F_{\tau}$.
The likelihood  $\chi^2$ is given by
\begin{equation}
 \chi^2 = \sum_{\tau,\tau',\omega}(K_{\tau',\omega}S_1(\omega) - < F_{\tau'}>)^T C^{-1}_{\tau',\tau} (K_{\tau,\omega}S_1(\omega) - <F_\tau>)
\end{equation}
where the kernel $K_{\tau,\omega}$ is defined as
\begin{equation}
K_{\tau,\omega} = \exp ( -\tau \omega) + \exp(-(\beta -
\omega)\tau)
\end{equation}
The covariance matrix, $C_{\tau,\tau'}$, describes the correlation
between the different time slices $\tau$ for a given atom
(``ring'' polymer). This matrix is defined as
\begin{equation}
C_{\tau,\tau'}= <{F_{\tau} F_{\tau'}}> -
<{F}_{\tau}><{F}_{\tau'}>,
\end{equation}
where $<{F_\tau}>$ is obtained as an average over all atoms at a
given time slice $\tau$. Because ${F_\tau}$ is periodic as one
goes around the polymer\cite{spect}, the summation  on $\tau$ is
done for $\tau = 1,M/2$, where $M$ is the total number of
time-slices in a ``polymer ring''.

The entropy term $S_{ent}$ is added to  $\chi^2$ in order to make
the reconstruction procedure better
conditioned\cite{Ceperley_RMP,MaxEnt}. We remark here that in some
QMC simulations only the diagonal elements of
$C{_{\tau,\tau'}}$ are taken into account~\cite{MaxEnt}, but here
we use all the elements, because the $<{F_\tau}>$ at different $\tau$
are correlated with each other.

Although $\chi^2$ measures how closely any form of
$S_1(\omega)$ approximates the solution of Eq. (7), one cannot
determine $S_1(\omega)$ reliably from PIMC using $\chi^2$
alone.\cite{Ceperley_RMP,MaxEnt} To make this determination, one needs to add the
entropy term $S_{ent}$ to $\chi^2$ in (9) to make the
reconstruction procedure better conditioned. The entropy term is
given by
\begin{equation}
S_{ent}(\omega) = -\int_{0}^{\infty} d\omega \left( S_1(\omega)\ln
\frac{S_1(\omega)}{m(\omega)} + S_1(\omega) - m(\omega) \right)
\end{equation}
where $m(\omega)$ includes our prior knowledge about the
properties of $S_1(\omega)$, examples of which were given above.
The simplest choice of $S_1(\omega)$ is the flat model, in which
$m(\omega)$=const for a selected range of frequencies and zero
otherwise. We took a cutoff frequency corresponding to an energy
of 100K. This flat model is used as an input for most  of
our simulations. In addition, we used a ``self-consistent'' model
where the output of a previous MaxEnt reconstruction is used as
input for the next MaxEnt reconstruction in an iterative
fashion~\cite{Ceperley_RMP,spect} The flat model is taken for the
initial iteration. Finally, we also tried $m(\omega)$  with
Gaussian and Lorentzian shape, with peaks given by the SCP theory
and experiment. As explained below, the outcome is not very
sensitive to the choice of $m(\omega)$ provided the PIMC data are
of good quality.

Finally, we discuss the parameter $\alpha$ in (9). The magnitude
of this parameter controls the relative weight of the PIMC data
vs. the entropy term in the determination of $S_1(\omega_1)$.
There are different strategies to obtain $\alpha$.  In our
simulations we used both the ``classical'' MaxEnt
method~\cite{MaxEnt} and random walk sampling.\cite{spect} The
``classical'' MaxEnt method picks the best value of $\alpha$,
while random walk sampling calculates a distribution of $\alpha$,
$\pi(\alpha)$. Next, for each value of $\alpha$ we calculate
$S_1(\omega)$. The final $S_1(\omega)$ is obtained as a weighted
average over $\alpha$. We found that when the collected PIMC data
is of good quality, the distribution $\pi(\alpha)$ becomes sharply
peaked and symmetric, and the phonon spectra obtained by means of
``classical'' MaxEnt and random walk are almost the same. Good
quality data are characterized by an absence of correlation between
sequential PIMC steps and by small statistical errors.
\subsection{\label{sec:level1} Our implementation }
The MaxEnt method assumes that the distribution of the sampled $
F_{\tau}$ is Gaussian.\cite{MaxEnt}  We re-block\cite{MaxEnt} the
sampled values of $ F_{\tau}$ in order to reduce the correlations
and to make the distribution as close as possible to a Gaussian,
with zero third (skewness) and fourth (kurtosis) moments.

The criterion determining the minimum number of sampled data
points comes from the properties of the covariance matrix
$C_{\tau,\tau'}$. If there are not enough blocks of data the
covariance matrix becomes pathological.\cite{MaxEnt} Therefore,
the number of blocks  must be larger than the number of time
slices in a ``ring''polymer. In our simulations, each atom is
represented by a ``ring''polymer with 64 time slices. We collected
at least 300 blocks in each simulation run.  We found that at
least 10000 data points were required in order to obtain the 300
blocks. Each simulation run took about two weeks of 12 Pentium III
PCs running in parallel.

Statistical errors were estimated by running the PIMC simulations
10 times, with different initial conditions in each case. After each run,
$S({\bf q }, \omega)$ was extracted using the MaxEnt method. The
phonon energy for a given ${\bf q}$ was then calculated by
averaging the positions of the peak of $S({\bf q }, \omega)$ over
the set of the simulation runs. The error bars of each point shown
in the figures below represent the standard deviation.
\begin{figure}% h - stands for here
\includegraphics{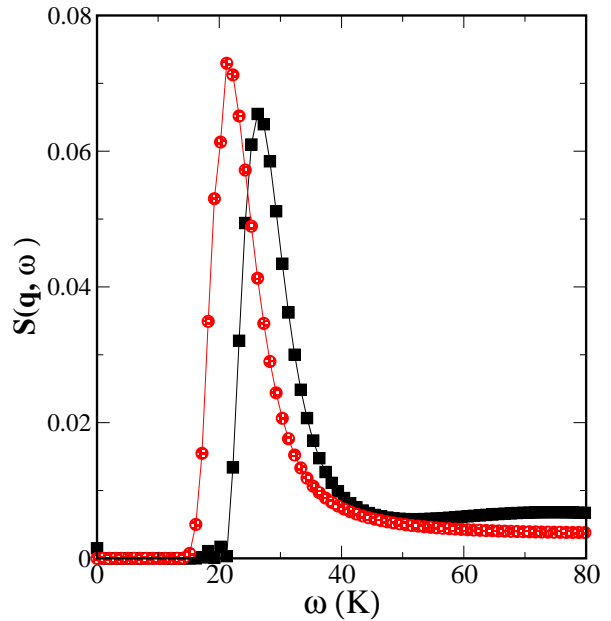}
\caption { \label{S1} (Color online) Longitudinal component of the dynamic
structure factor, $S({\bf q }, \omega)$,  for q = 0.83 r.l.u. along the [100] direction:
single phonon contribution (circles, red online), and total structure factor
(squares, black online). The lines are a guide to the eye.}

\end{figure}

In the simulations we used samples containing between 128 and 432
atoms. This allowed us to calculate $S({\bf q }, \omega)$ for
values of $q$ between 0.17 and 1 in relative lattice units
(r.l.u.= 2$\pi/a$, where $a$ is the lattice parameter). The number
density was set to $\rho = 0.02854~(1/A^3)$ and the temperature
was $T=1.6~K$. A perfect bcc lattice was chosen as the initial
configuration. The effects
of Bose statistics are not taken into account
in our simulation,
which is a reasonable approximation for the solid phase.
 A typical example of the calculated dynamic
structure factor is shown in Fig.\ref{S1}. The figure shows both
the single phonon contribution $S_1({\bf q },\omega)$ and the
total $S({\bf q },\omega)$ for a longitudinal phonon along the
[001] direction. To illustrate the difference between $S_1({\bf q
},\omega)$ and $S({\bf q }, \omega)$, we chose to show the results
for $\bf q$ close to the boundary of the Brillouin zone. This
difference is discussed below.

\section{\label{sec:level1}  Results }
\subsection{\label{sec:level1} Phonon spectra }
\begin{figure} [h]% h - stands for here
\includegraphics{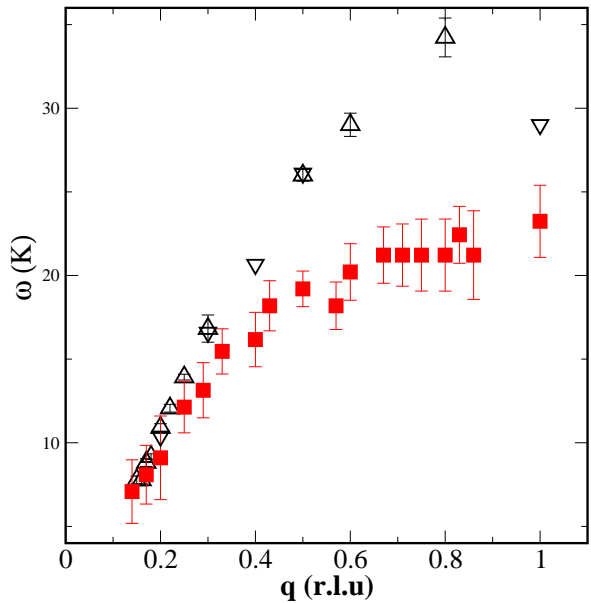}
\caption { \label{l100} (Color online) Calculated dispersion relation of the
L[001] phonon branch (squares, red online) using $S_1({\bf q },\omega)$.
Experimental data are from \cite{Osgood,Osgood2} (triangles up)
and from \cite{Tuvi} (triangles down). The error bars represent statistical uncertainty.}
\end{figure}

\begin{figure} [h]% h - stands for here
\includegraphics{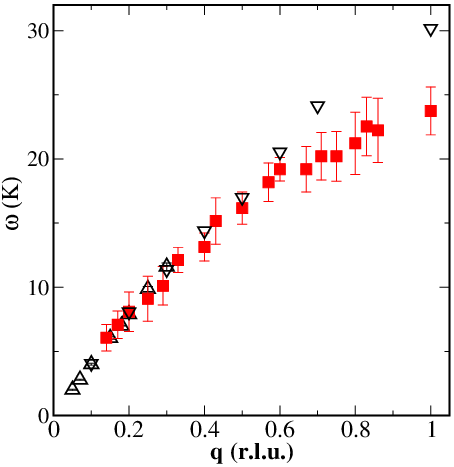}
\caption { \label{t001} (Color online) Calculated dispersion relation of the
T[001] phonon branch (squares,  red online) using $S_1({\bf q },\omega)$.
Experimental data are from \cite{Osgood,Osgood2} (triangles up)
and from \cite{Tuvi} (triangles down). The error bars represent statistical uncertainty.}
\end{figure}

\begin{figure} [h]% h - stands for here
\includegraphics{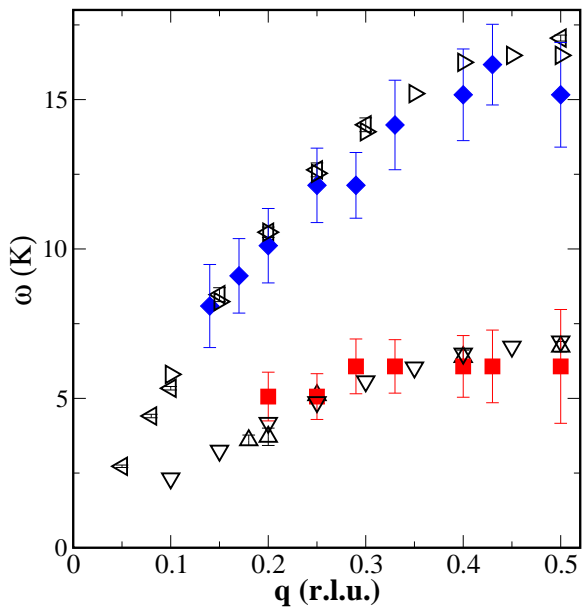}
\caption { \label{t011} (Color online) Calculated  dispersion relations of
transverse phonon branches along [011] using $S_1({\bf q
},\omega)$. Calculated values are shown for the $T_1$ branch
(squares, red online) and $T_2$  branch (diamonds, blue online). Experimental data are from
\cite{Osgood,Osgood2} ( $T_1$-triangles up, $T_2$-triangles left)
and from \cite{Tuvi} ($T_1$-triangles down, $T_2$-triangles
right). The error bars represent statistical uncertainty.}
\end{figure}

\begin{figure} [h]% h - stands for here
\includegraphics{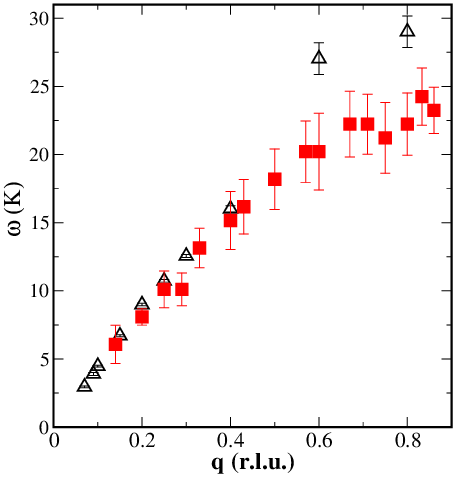}
\caption { \label{t111} (Color online) Calculated dispersion relation of the
T[111] phonon branch (squares, red online) using $S_1({\bf q },\omega)$.
Experimental data are from \cite{Osgood,Osgood2} (triangles up)
and from \cite{Tuvi} (triangles down).
The error bars represent statistical uncertainty.}
\end{figure}

The calculated longitudinal and transverse phonon spectra of solid
$^4He$ in the bcc phase along the main crystal directions ([001],
[111] and [011]) are shown in Figs.~\ref{l100} - \ref{t111}.
We compare our results with the experimental data measured by
inelastic neutron scattering from bcc $^4\text He$ with a molar
volume of 21.1 $\text cm^3$ at T = 1.6 K., by Osgood et
al,\cite{Osgood,Osgood1,Osgood2} and by Markovitch et
al.~\cite{Tuvi}
\begin{figure} [h]% h - stands for here
\includegraphics{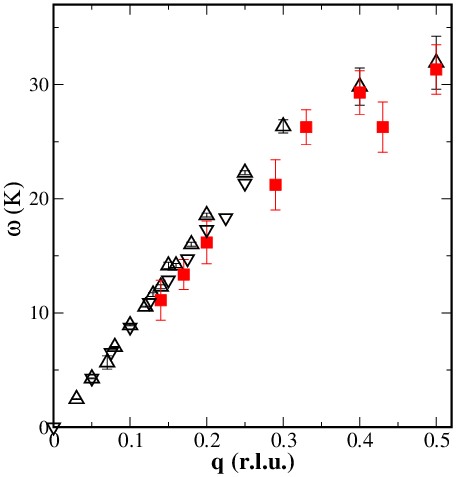}
\caption { \label{l111}(Color online) Calculated dispersion relation of the
L[011] phonon branch (squares, red online) using $S_1({\bf q },\omega)$.
Experimental data are from \cite{Osgood,Osgood2} (triangles up)
and from \cite{Tuvi} (triangles down). The error bars represent statistical uncertainty.}
\end{figure}
\begin{figure} [h]% h - stands for here
\includegraphics{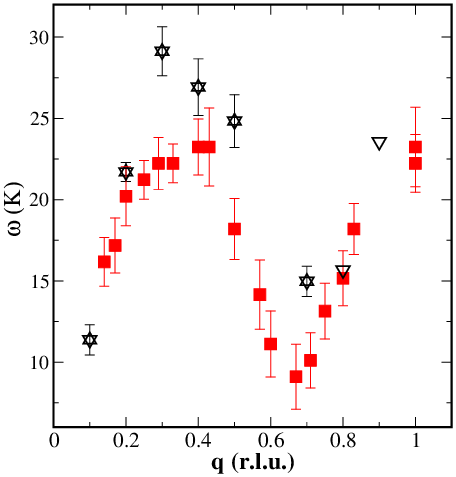}
\caption { \label{l111} (Color online) Calculated dispersion relation of the
L[111] phonon branch (squares, red online) using $S_1({\bf q },\omega)$.
Experimental data are from \cite{Osgood,Osgood2} (triangles up)
and from \cite{Tuvi} (triangles down).The error bars represent statistical uncertainty.}
\end{figure}
\begin{figure} [h]% h - stands for here
\includegraphics{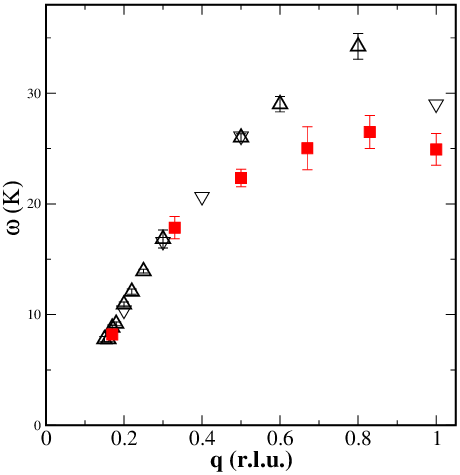}
\caption { \label{lt001} (Color online) Calculated dispersion
relation of the L[001] phonon branch (squares, red online) using
$S({\bf q},\omega)$. Experimental data are from
\cite{Osgood,Osgood2} (triangles up) and from\cite{Tuvi}(triangles
down). The error bars represent statistical uncertainty.}
\end{figure}
\begin{figure} [h]% h - stands for here
\includegraphics{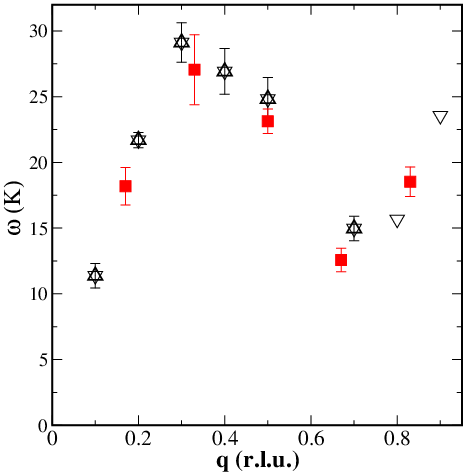}
\caption { \label{lt111} (Color online) Calculated dispersion
relation of the L[111] phonon branch (squares, red online) using
$S({\bf q},\omega)$. Experimental data are from
\cite{Osgood,Osgood2} (triangles up) and from
\cite{Tuvi}(triangles down). 
The error bars represent statistical uncertainty. }
\end{figure}
\begin{figure} [h]%  - stands for here
\includegraphics{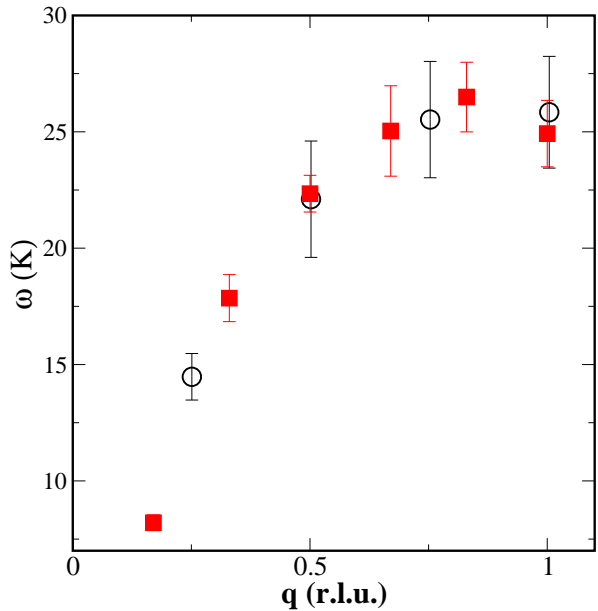}
\caption { \label{cmpr}  (Color online) A comparison of dispersion
relations of the L[001] phonon branch  obtained in the present
work using $S({\bf q},\omega)$(squares, red online), with the same
relation calculated by means of the Shadow Wave Function
technique\cite{Reatto} (circles). The error bars represent
statistical uncertainty.}
\end{figure}

As expected, the agreement between our simulations of $S_1({\bf
q},\omega)$ and experiment is very good at small ${\bf q}$, where
one-phonon excitation is  the most significant contribution to
$S({\bf q},\omega)$. As ${\bf q}$ increases, higher order
processes become significant, and the calculated values deviate
from the experimental data, especially along [001] and [111]. In
the case of longitudinal phonons, it is possible to calculate
their energies using the total $S({\bf q},\omega)$ obtained
directly from Eq. (1) instead of just the single phonon
contribution. The dispersion relations calculated with $S({\bf
q},\omega)$ are shown in Figs.~\ref{lt001} and \ref{lt111}. It is
evident that using the total scattering function improves the
agreement with experiment at large ${\bf q}$, especially for the
[111] direction.

We point out that the calculated phonon branches $T_1(110),
T_2(110)$ and $L(110)$ show good fit to the experimental data. Our
results were obtained with the two body potential, which takes the
He atoms as point particles. Gov et al.~\cite{Gov} suggested that
one needs to go beyond this approximation to obtain the $T_1(110)$
phonon branch in good agreement with experiment. Gov's approach
also predicts the new excitation branch observed
recently\cite{Emil}. Although the calculated $T_1(110)$ branch is
in agreement with experiment without any additional assumptions,
we were not able to see the new excitation in our simulations.
Experimentally, this excitation is about an order of magnitude
less intense than a phonon. It is best observed in scattering
experiments done with very small $\bf q \leq$ 0.1
r.l.u\cite{Tuvi}. Both these factors make it very difficult to
search for this excitation in simulations. Whether it can be found
in this approach remains an open question.

In addition to experimental results, our simulations can  also be
compared with those of Galli and Reatto\cite{Reatto}, who used the
Shadow Wave Function (SWF) approach to calculate the longitudinal phonon
branches of bcc $^4$He. As shown in Fig.\ref{cmpr},
the overall agreement between these PIMC simulations
and SWF results is good.

\subsection{\label{sec:level1} Vacancies }
 Recent experimental work
\cite{Kim} revived the interest in point defects, such as
vacancies. It is therefore interesting to examine the influence of
vacancies on the properties of the solid. We repeated our
calculation of the phonon branches in the presence of 0.23$\%$
vacancies (1 atom of 432). Within the statistical error bars, we
found no difference between the phonon energies with or without
vacancies. Galli and Reatto\cite{Reatto} found that vacancies
lower the energies of the phonons close to the boundary of the
Brillouin zone. However, in their simulation they used a
concentration of vacancies of 0.8$\%$, so that the cumulative
effect may be larger. We also calculated the vacancy
formation energy, $\Delta E_{v}$, according to Pederiva et
al.~\cite{Pederiva}
\begin{equation}
\label{evas}
\Delta E_{v} = (E(N-1,\rho) - E(N,\rho))(N-1),
\end{equation}
where $E(N,\rho)$ is the total energy of $N$ atoms. The energy
$E(N,\rho))$ was calculated for a perfect crystal, while
$E(N-1,\rho)$ was calculated after removing one atom. The density
of two systems was kept the same by adjusting the lattice
parameter.  Values of $\Delta E_{v}$  calculated using the PIMC,
Shadow Wave Function (SWF) \cite{Chester} and Shadow Path Integral
Ground State (SPIGS)\cite{Reatto} methods are summarized in
Table~\ref{tab1}. In addition, we calculated $\Delta E_{v}$ at 
constant volume, which is the condition usually realized in
experiments rather than constant density. We obtained $\Delta
E_{v}$= 5.7 $\pm$ 0.7 K. The lower value arises since the
repulsive part of the potential is weaker in a sample having 
lower density. There is no generally accepted experimental
value~\cite{Burns} of $\Delta E_{v}$. According to NMR
studies\cite{Allen, Schuster} the energy of vacancy formation in
the bcc phase is $\Delta E_{v}=6.5\pm 0.2 $(K), while X-ray
studies\cite{Fraas} suggest that $\Delta E_{v}=9\pm 1 $(K). We
comment here that the calculated values of $\Delta E_{v}$ are
significantly smaller than 14K, the energy of the new excitation
observed by Markovitch et al.\cite{Emil,Tuvi}. Hence, this new
excitation does not seem to be a simple vacancy.
\begin{table}

\caption{\label{tab1} Calculated energy of formation of a vacancy,
$\Delta E_{v}$, for bcc solid $^4He$. N is the number of atoms
used in each of the simulations.}
\begin{tabular}{|c|c|c|c|c|} \hline
source             & method & density (1/$A^3$) & N & $\Delta E_{v}$~(K)   \\ \hline \hline
this work          & PIMC   & 0.02854 &   128 & $ 10.57 \pm 0.38 $  \\ \hline
this work          & PIMC   & 0.02854 &   250 & $ 9.96  \pm  0.89 $  \\ \hline
ref.\cite{Chester} & SWF    & 0.02854 &   128 & $ 8.08  \pm 2.76  $  \\ \hline
ref.\cite{Chester} & SWF    & 0.02854 &   250 & $ 6.69 \pm 3.86  $  \\ \hline
ref.\cite{Reatto}  & SWF    & 0.02898 &   128 & $ 8.9  \pm 0.3    $  \\ \hline
ref.\cite{Reatto}  & SPIGS  & 0.02898 &   128 & $ 8.0  \pm 1.3    $  \\ \hline

\end{tabular}
\end{table}

\section{\label{sec:level1}  Conclusions }
We calculated the dynamic structure factor for solid helium in the
bcc phase using PIMC simulations and the MaxEnt method. PIMC
was used to calculate the intermediate scattering function in the
imaginary time from which the dynamic structure factor was
inferred with the MaxEnt method. We extracted the longitudinal and
transverse phonon branches from the one-phonon dynamic structure
factor. To the best of our knowledge this is the first simulation
undertaken for the transverse
branches.  At small ${\bf q}$, where the one-phonon excitation is
the most significant contribution to the dynamic structure factor,
the agreement between our simulations and experiment is very good.
At large ${\bf q}$, multi-phonon scattering and interference
effects~\cite{Glyde} becomes important. Consequently, the position
of the peak of in the $S_1({\bf q},\omega)$ does not correspond to
the position of the peak in the $S({\bf q},\omega)$, and the
phonon energies calculated from $S_1({\bf q},\omega)$ are too low.
If $S({\bf q},\omega)$ is used instead of $S_1({\bf q},\omega)$,
the agreement with experiment is significantly improved. We
repeated the simulations in the presence of 0.23$\%$ of vacancies,
and found no significant differences in the phonon dispersion
relations. We also calculated the formation energy of a vacancy 
 both at constant density and at a constant volume.

\begin{acknowledgments}
We wish to thank D. Ceperley for many helpful discussions and for
providing us with his  UPI9CD PIMC code. We are grateful to  N. Gov, O.
Pelleg and S. Meni for discussions. This study
was supported in part by the Israel Science Foundation and by the
Technion VPR fund for promotion of research.
\end{acknowledgments}


\begin{thebibliography}{13}
{
\bibitem{Glyde}
H. R. Glyde , {\it Excitations in Liquid and Solid Helium} , Clarendon Press, Oxford, (1994).
\bibitem{Horner}
H. Horner,  J. Low. Temp. Phys. {\bf 8}, 511, (1972).
\bibitem{Gov}
N. Gov and E. Polturak, Phys. Rev.  {\bf B }, {\bf 60}, 1019, (1999).
\bibitem{spect}
M. Boninsegni and D. M. Ceperley, J. Low. Temp. Phys.  {\bf 104},
336, (1996).
\bibitem{Reatto}
D. E. Galli and L. Reatto,  J. Low. Temp. Phys., {\bf 134}, 121,
(2004).
\bibitem{Kim}
E. Kim and M. H. W. Chan, Science,{\bf 305}, 1941, (2004).
\bibitem{Emil}
T. Markovich, E. Polturak, J. Bossy, and E. Farhi, Phys. Rev.
Lett., {\bf 88}, 195301, (2002).
\bibitem{Tuvi}
T. Markovich, {\it Inelastic-Neutron Scattering from bcc $^4$He},
PhD. Thesis, Haifa, Technion, (2001).
\bibitem{Chester}
B. Chaudhuri, F. Pederiva, and G. V. Chester, Phys. Rev. B., {\bf
60}, 3271, (1999).
\bibitem{Ceperley_RMP}
D. M. Ceperley, Rev. Mod. Phys., {\bf 67}, 279, (1995).

\bibitem{Book}
K. Ohno, K. Esfarjam, Y. Kawazoe,  {\it Computational Material
Science; From Ab Initio to Monte Carlo Methods}, Springer, Berlin,
1999.
\bibitem{Bernu}
B. Bernu and D. M. Ceperley, {\it Quantum Simulations of
Complex Many-Body Systems: From Theory to Algorithms }, NIC Series {\bf 10}, Julich, 2002.
\bibitem{Aziz}
R. A. Aziz, A. R. Janzen, and M. R. Moldover, Phys. Rev. Let. {\bf 74},  1586 (1995).
\bibitem{Born}
M. Born and K. Huang, {\it Dynamical theory of crystal lattices},
Clarendon Press,  Oxford, (1954).
\bibitem{Bruesch}
P. Bruesch,{\it Phonons, theory and experiments I : lattice
dynamics and models of inter atomic forces} , Berlin, Springer,
(1982).
\bibitem{MaxEnt}
M. Jarrell and J. E. Gubernatis, Phys. Rep.,  {\bf 269}, 133, (1996).
\bibitem{Silver}
J. E. Gubernatis, M. Jarrell, R. N. Silver, and D. S. Sivia
Phys. Rev. B {\bf 44}, 6011, (1991).
\bibitem{Osgood}
E. B. Osgood, V. J. Minkiewicz, T. A. Kitchens, and G. Shirane, Phys. Rev. A  {\bf 5}, 1537 (1972).
\bibitem{Osgood1}
E. B. Osgood, V. J. Minkiewicz, T. A. Kitchens, and G. Shirane, Phys. Rev. A  {\bf 6}, 526 (1972).
\bibitem{Osgood2}
V. J. Minkiewicz, T. A. Kitchens, G. Shirane, and E. B. Osgood, Phys. Rev. A {\bf 8}, 1513 (1973).
\bibitem{Pederiva}
F. Pederiva, G. V. Chester, S. Fantoni, and L. Reatto, Phys. Rev.
B {\bf 56}, 5909, (1997).
\bibitem{Burns}
C. A. Burns and J. M. Goodkind, J. Low. Temp. Phys., {\bf 95},
695, (1994).
\bibitem{Allen} A. R. Allen, M. G. Richards, and G. Sharter, J. Low. Temp. Phys., {\bf 47}, 289,
(1982).
\bibitem{Shuster} I. Schuster, E. Polturak, Y. Swirsky, E. J. Schmidt, and S. G. Lipson, J. Low Temp. Phys.
{\bf 103}, 159, (1996).
\bibitem{Fraas}
B. A. Fraass, P. R. Granfors, and R. O. Simmons, Phys. Rev. B., {\bf 39}, 124, (1989).


}
\end{thebibliography}
\end{document}